\documentclass{article}

\usepackage{arxiv}
\usepackage[utf8]{inputenc} 
\usepackage[T1]{fontenc}    
\usepackage{hyperref}       
\usepackage{url}            
\usepackage{booktabs}       
\usepackage{amsfonts}       
\usepackage{nicefrac}       
\usepackage{microtype}      
\usepackage{lipsum}
\usepackage{amsmath}
\usepackage{graphicx}
\usepackage{subfigure}
\usepackage{tabularx}
\usepackage{array}
\usepackage{multirow}
\usepackage{caption}
\usepackage[title,toc,titletoc,header]{appendix}

\usepackage{hyperref}
\hypersetup{
 colorlinks=true,
 linkcolor = blue,
 filecolor = blue, 
 urlcolor = blue,
 citecolor = blue
}

\title{Information Mining for COVID-19 Research From a Large Volume of Scientific Literature}

\author{
  Sabber Ahamed \\
  Asurion\\
 Nashville, TN 38152 \\
  \texttt{sabbers@gmail.com} \\
  \and
   {\bf Manar D. Samad} \\
  Department of Computer Science\\
 Tennessee State University\\
 Nashville, TN 37209 \\
  \texttt{msamad@tnstate.edu} \\
}

\begin{document}
\maketitle

\begin{abstract}

The year 2020 has seen an unprecedented COVID-19 pandemic due to the outbreak of a novel strain of coronavirus in 180 countries. In a desperate effort to discover new drugs and vaccines for COVID-19, many scientists are working around the clock. Their valuable time and effort may benefit from computer-based mining of a large volume of health science literature that is a treasure trove of information. In this paper, we have developed a graph-based model using abstracts of 10,683 scientific articles to find key information on three topics: transmission, drug types, and genome research related to coronavirus. A subgraph is built for each of the three topics to extract more topic-focused information. Within each subgraph, we use a betweenness centrality measurement to rank order the importance of keywords related to drugs, diseases, pathogens, hosts of pathogens, and biomolecules. The results reveal intriguing information about antiviral drugs (Chloroquine, Amantadine, Dexamethasone), pathogen-hosts (pigs, bats, macaque,  cynomolgus), viral pathogens (zika, dengue, malaria, and several viruses in the $coronaviridae$ virus family), and proteins and therapeutic mechanisms (oligonucleotide, interferon, glycoprotein) in connection with the core topic of coronavirus. The categorical summary of these keywords and topics may be a useful reference to expedite and recommend new and alternative directions for COVID-19 research.

\end{abstract}

\keywords{coronavirus \and COVID-19 \and text mining \and graph theory \and topic modeling \and betweenness centrality} 

\section{Introduction}
Severe acute respiratory syndrome (SARS) emerged in 2002 following the outbreak of a new SARS-related coronavirus (SCV), which has resulted in hundreds of deaths world-wide after infecting thousands of individuals~\citep{Fouchier2003}. Since then, there has been a plethora of research aiming at understanding the etiology of the disease~\citep{Weiss2005}, developing more reliable and faster diagnostic solutions~\citep{Yamaoka2016}, and innovating antiviral drugs~\citep{Keyaerts2009} and vaccines~\citep{Yong2019}. Many of these studies have successfully repurposed previous findings to solve new challenges in biology and medicine. For example, if a group of viruses shares a common protein structure, then therapies for one viral infection can be repurposed for new diseases like COVID-19. Unfortunately, viruses often mutate to several challenging strains that are capable of infecting humans in a highly contagious manner. This eventually triggers the onset of zoonotic infections, i.e., animal-to-human and human-to-human transmission of the virus strain~\citep{Ozawa2013}. 

The ongoing global pandemic of coronavirus disease (COVID-19) has emerged due to a new strain of SCV named as the SARS coronavirus 2 (SARS-CoV-2)~\citep{Zhu2020} that is arguably hosted by bats~\citep{Wu2020}. As of March 31st, 2020, 857,487 people of 180 countries have been infected by the virus in about three months, taking a toll of 42,107 human lives as the numbers continue to grow exponentially~\citep{Dong2020}. This outbreak has already started to cause an unprecedented impact on humankind at a scale that threatens to reshape the global economy, human lives, and the world beyond. Unfortunately, it entails substantial time and effort to study these new strains of viruses before developing a targeted and fail-safe drug or vaccine. Without any immediate solution, several countries have recently approved the use of hydroxychloroquine to treat critically ill patients diagnosed with COVID-19 despite its known side-effects~\citep{Liu2020, Masuelli2017}. Notably, hydroxychloroquine is a drug for treating patients diagnosed with malaria~\citep{ben2012}. This repurposing of malaria drug to treat coronavirus related patients suggests that topics of virology, genetic engineering, bio-informatics, pharmacology in the literature may have valuable recommendations for COVID-19 research.  

Since the 2002 outbreak of coronavirus, many scholarly articles have been published on the topics of SARS and SCV. It is not feasible for scientists to survey such a massive volume of diverse literature to derive useful cross-disciplinary insights within a short period of time. Computer-based modeling of text documents can be a powerful approach in search of valuable information from a large collection of diverse literature. In this paper, we have developed a graph-based model using a large volume of scientific literature to discover latent keywords that may recommend new information to combat the battle against COVID-19. 

\section{Technical Background}

In text analytics, keywords searching plays an important role in document summarization~\citep{Hernandez-castaneda2020}, classification~\citep{Diaz-Manriquez2018}, and topic modeling~\citep{Jelodar2019}. However, the task of preserving semantic and syntactic information of a language in numerical models is not trivial. There are three broad categories of text-to-numeric representation models. First, term-frequency-inverse-document frequency (TF-IDF)~\citep{Yun-tao2005}, is one of the earliest methods for text representation based on word count. However, such models do not capture semantic relevance and syntactic ordering of the words. Second, manifold learning techniques, such as skip gram~\citep{Cheng2006} and GLOVe~\citep{Pennington2014}, have been developed to efficiently capture semantic relevance between words in numeric data vectors. These numeric embeddings do not directly map the intricate relationship between words within a document.  Third, graphs can be used to develop an interconnected network of words of a document to directly model the syntactic ordering and connectivity between words. A graph-based keyword extraction model using collective node weight (KECNW) has been proposed based on node-edge rank centrality measure~\citep{Biswas2018}.  Several attributes, such as the distance from the central node, the selectivity of each node based on a number of its connecting edges, and the importance of neighboring nodes are taken into account to determine the weight of each node. The node weights are used to rank individual words for keyword selection. Additionally, graph-based models provide informative and interactive visualization of complex interconnected network of words in a document~\citep{Rusu2009}. The goal of this paper is to develop a dense graphical network of words to summarize thousands of research articles into topics related to coronavirus. The collection of influential nodes or keywords may provoke new research topics for the ongoing COVID-19 research.

\section{Methodology}
The steps involved in the proposed research methodology are discussed below.  
\subsection{Dataset}

In this work, we use the publicly available COVID-19 Open Research Dataset (CORD-19) released by a collaborative initiative of the Allen Institute for AI, Chan Zuckerberg Initiative (CZI), Georgetown University's Center for Security and Emerging Technology (CSET), Microsoft, and the National Library of Medicine (NLM) at the National Institutes of Health (NIH) at the request of the White House Office of Science and Technology Policy. The dataset includes about 13,000 scholarly articles with full texts in the areas of life science, including pharmacology, immunology, physiology, microbiology, clinical medicine, genetics, veterinary, and many more. The articles are related to the study of COVID-19, SARS-CoV-2, and similar other coronaviruses. The dataset comes with metadata and actual article texts in structured data fields that identify title, paper and publication information, abstract, and the main body texts. More details of the dataset can be found on the Kaggle website (\url{https://www.kaggle.com/allen-institute-for-ai/CORD-19-research-challenge}), which is a community-based data science platform for solving problems around the world.

\subsection {Proposed Graph Model}

\begin{figure}  [t]
\centering
\subfigure {\includegraphics[trim=0cm 0cm 0cm 0cm, clip=true, totalheight=0.06\textheight, angle=0 ]{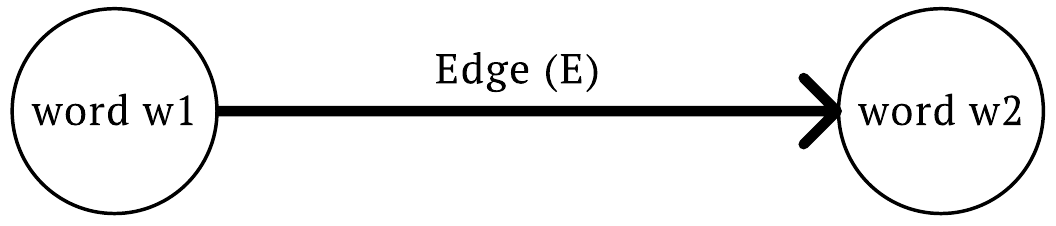}}
\subfigure {\includegraphics[trim=0cm 0cm 0cm 0cm, clip=true, totalheight=0.22\textheight, angle=0 ]{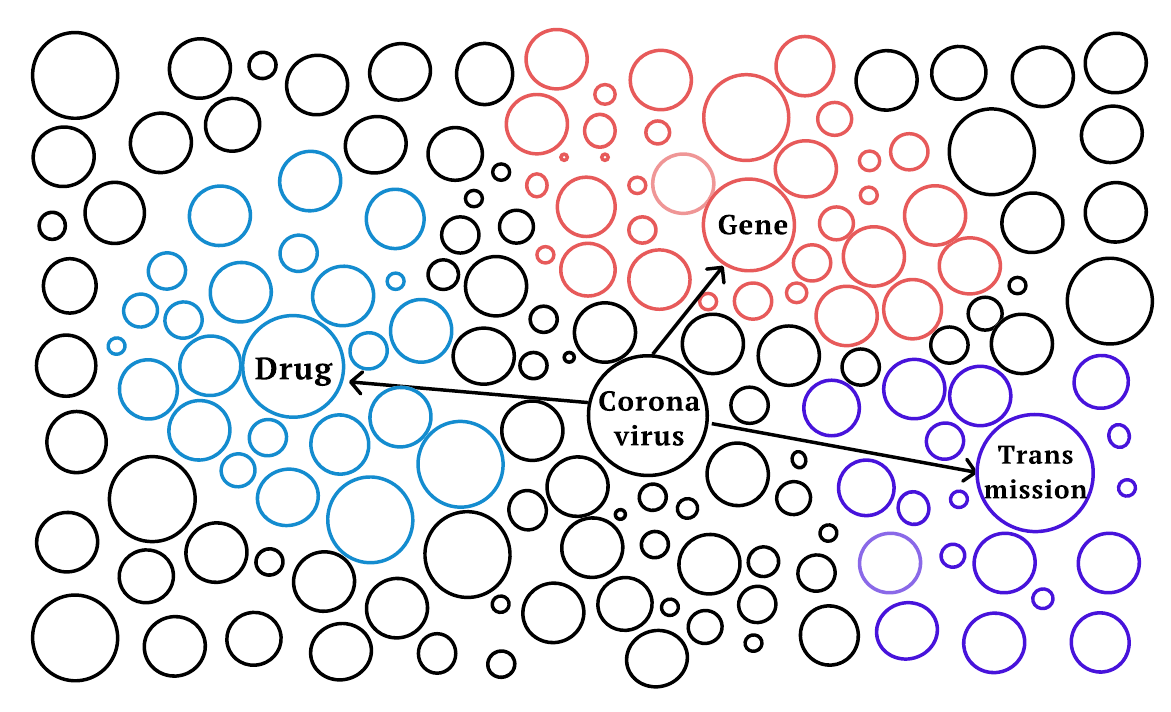}}
\caption{Left: A pair of graph nodes with a directed edge between words w1 and w2. Right: A schematic diagram to illustrate the central $coronavirus$ node. The size of neighoboring nodes varies with BC measurements of their corresponding words. Three influential nodes with high BC measurements and their subgraphs are color coded.}
\label{graph01}
 \end{figure}

We propose a directed graph $G(N, E)$ based model that is represented by a set of nodes $N$ and a set of edges $E$ with directions. Each node $n \in N$ represents a unique word. A directed edge $e \in E$ connects two adjacent words based on their positioning in a sentence. The direction of an edge points from the preceding word to the succeeding word in a sentence as shown in Figure~\ref{graph01}. Using this principle, a sequence-based text graph is first generated using texts of the articles. In the next step, we determine the importance of each node in the graph based on the concept of 'centrality'. In general, a node with a higher centrality has more central positioning in the graph. For example, the node with relatively shortest paths to all other nodes in the graph has the highest measure of centrality~\citep{Hage1995}. In this paper, we measure the centrality of a node using the betweenness centrality (BC) algorithm that is widely applied in social network analysis. However, the BC algorithm has not been well studied in topic selection and modeling with a large volume of texts, as we propose in this article. The BC algorithm is originally proposed by~\citet{freeman1977set} and later upgraded by~\citet{Brandes2001} to improve its computational performance.

In BC measurement, the node with highest centrality will be located in the highest number of shortest paths between all other node pairs in the graph. In other words, if a word frequently appears between the shortest paths of other word pairs, then the word will have a high value of centrality. Thus, the BC algorithm finds influential words that are imperative for understanding the relationship between word pairs. Such transitive words are weighted as representative nodes in the graph and selected as key topics for subsequent analysis. Intuitively, this approach is more sophisticated and informative than using a distance-based  centrality measurement. If (u, v, w)$\in$ N such that $u \neq v \neq w$, the shortest path between any two nodes can be calculated using a standard Dijkstra's algorithm. To measure BC of a node $w$ in the graph, the number of shortest paths between all other node pairs is first obtained and denoted by $\sigma_{uv}$. Next, the number of shortest paths ($\sigma_{uv}(w)$) is counted between all other node pairs that include the target node $w$ on the path. The ratio of these two counting numbers is calculated for each pair of nodes as the pair-dependency ($\delta_{u, v}$) below.
\begin{eqnarray}
\delta_{u, v} = \frac{\sigma_{uv} (w)}{\sigma_{uv}}.
\end{eqnarray}
The sum of pair-dependency across all possible pairs (excluding $w$) in the graph yields the BC of the target node $w$ as below.
 \begin{eqnarray}
BC (w) = \sum_{u \neq v \neq w} \frac{\sigma_{uv} (w)}{\sigma_{uv}}.
\end{eqnarray}

Each node representing a unique word can be a representative topic by itself. The corpus words are ranked by BC measurements to select semantically important key topics with relatively high BC values. The key topics are then used to create their corresponding subgraphs. Each topic word is the center of the subgraph, and its neighboring words are the constituents of the topic. The proposed method does not require the number of topics in advance compared to other topic modeling algorithms such as latent Dirichlet allocation (LDA)~\citep{NIPS2010_3902} and latent semantic analysis (LSA)~\citep{Landauer1998}.  

\subsection{Data Processing}
We develop the proposed model and perform the text analysis on a Python programming environment using several Python packages. These packages include natural language tool kit (\texttt{NLTK}) for basic text analysis and regular expressions package (\texttt{re}) for text preprocessing. A graph database \texttt{Neo4J} and a Python to database connector library  (\texttt{py2neo}) are used to build the graph. The first step in text processing is to identify and remove 'stop words' (words with little or no merit as keywords) from the corpus. This step not only reduces the complexity in the graph modeling but also puts more weights on relevant candidate words in the graph. The dataset includes research articles composed in Chinese, Russian, Arabic, and Korean languages. We confine our analysis to English articles by excluding all non-English words using regular expression (regex). Regex is also used to filter out undesired symbols and numeric data. The graph data, including nodes and edges, are exported to Gephi~\citep{Bastian2009} for better visualization and manipulation of the graph models and model-driven results. All the notebooks with source code and data are publicly shared in a Github project repository~(\url{https://github.com/msahamed/covid19-text-network}).

\subsection{Modeling and Analysis}
The proposed graph model is developed using all available article abstracts in the dataset. We assume that an abstract has succinct and sufficient information to represent the overall content of a research article. Once the global graph is developed using the article abstracts, we find several influential words based on their BC values that are linked to $coronavirus$. Next, we build separate subgraph for each of the selected key topic as shown in Figure~\ref{graph01}. Within each subgraph, neighboring nodes of a key topic are expected to reveal a variety of information about the topic word. The neighboring words are further grouped into several categories. Within each category, we rank order the neighboring words using BC values obtained from the subgraph model to identify their relative importance. We discuss our findings of this approach in the next section.

\section{Results and Discussion}

The CORD-19 dataset has a substantial number of articles with missing abstracts. Therefore, we use available abstracts for 10,683 research articles to build our proposed graph model. In the data processing step, we have created a list of 2796 stop words based on the word frequency and visual scrutiny to exclude them from the corpus. We first find the word $coronavirus$ in the global graph to select its three neighboring nodes with high BC values. Since BC values can be exponentially very large, we take natural logarithm of these values (log BC). We select three key topics: $transmission$, $drug$, and $gene$ with log BC values of 16.65, 16.75, and 18.27, respectively. In the following subsections, we discuss our findings on the three subgraph models corresponding to $transmission$, $drug$, and $gene$ topics, respectively.

 \begin{table}[t]
\centering
 \caption{Neighboring words in the $transmission$ subgraph linked to $coronavirus$.}
 \label{tab:trans_keywords}
 \centering
\begin{tabularx}{\textwidth}{l|Xc} 
\toprule
Topic & Connected nodes in the $transmission$ subgraph \\
\midrule
 Host & 'Rats', 'Livestock', 'Mammals', 'Goats', 'Camels', 'Mosquito', 'Pigs', 'Monkeys', 'Dogs', 'Poultry', 'Rodents', 'Horses', 'Bird', 'Chimpanzee', 'Honeybees', 'Mites', 'Puma', 'Snails', 'Infants', 'Ants', 'Flies' , Rodents, Drosophila\\ \midrule
Disease & 'Malaria', 'Rabies', 'Dengue', 'Chikungunya', 'Measles', 'Schistosomiasis', 'Monkeypox', 'Poliovirus', 'Cholera', 'Lyssavirus', 'Tuberculosis', 'Rhinovirus', 'Adenovirus', 'HFMD', 'HIV', 'Paramyxovirus', 'Zikavirus', 'Fungal', 'Leishmania', 'Myocarditis', 'Venereal'\\ \midrule
 Miscellaneous &  'Distance', 'Fever', 'Traveler', 'Contacts', 'Exposures', 'Humidity',  'Infants',  'Pregnancy', 'Interspecies', 'Mucosal', 'Reticulum', 'Zoonosis', 'Sanitization', 'Bioaerosols', 'Aerobiology', 'Tropisms', 'Human-to-human', 'Transspecies', 'Touches', 'MDRO', 'Fluoroquinolone'.\\
 \bottomrule
 \end{tabularx}
\end{table}

\subsection{Transmission Subgraph}
Studies related to infectious disease transmission are important in understanding the spread and growth of pathogens in different media. Table~\ref{tab:trans_keywords} shows the neighboring words of the $transmission$ subgraph. The subgraph reveals a number of host-pathogen interactions. Animal hosts such as poultry, camel, dogs, pigs, monkeys, chimpanzees may be candidates for studying the transmission mechanism of different zoonotic pathogens related coronavirus. The coronavirus related topic $transmission$ also connects to a number of diseases and pathogens, including rabies, dengue, tuberculosis, adenovirus, malaria, measles, and hand-foot-and-mouth disease (HFMD). These pathogens may have interesting traits in common with the transmission of coronavirus. Keywords such as multi-drug resistant organism (MDRO), tropism (the growth or movement of organisms), zoonosis (the topic of disease transmission from animals to humans), sanitization, bioaerosols have expectedly appeared in the subgraph of transmission topic. Additionally, the subgraph includes words related to different transmission media such as blood, respiratory, lung, water, aerosol, oronasal, gut, mouth, trachea, glands, droplet, saliva, airborne, and semen. Other miscellaneous keywords are presented in Table~\ref{tab:trans_keywords} that includes infants and pregnancy that are important topics for transmission studies.

 \begin{figure}[t]
\centering
\includegraphics[scale=0.65]{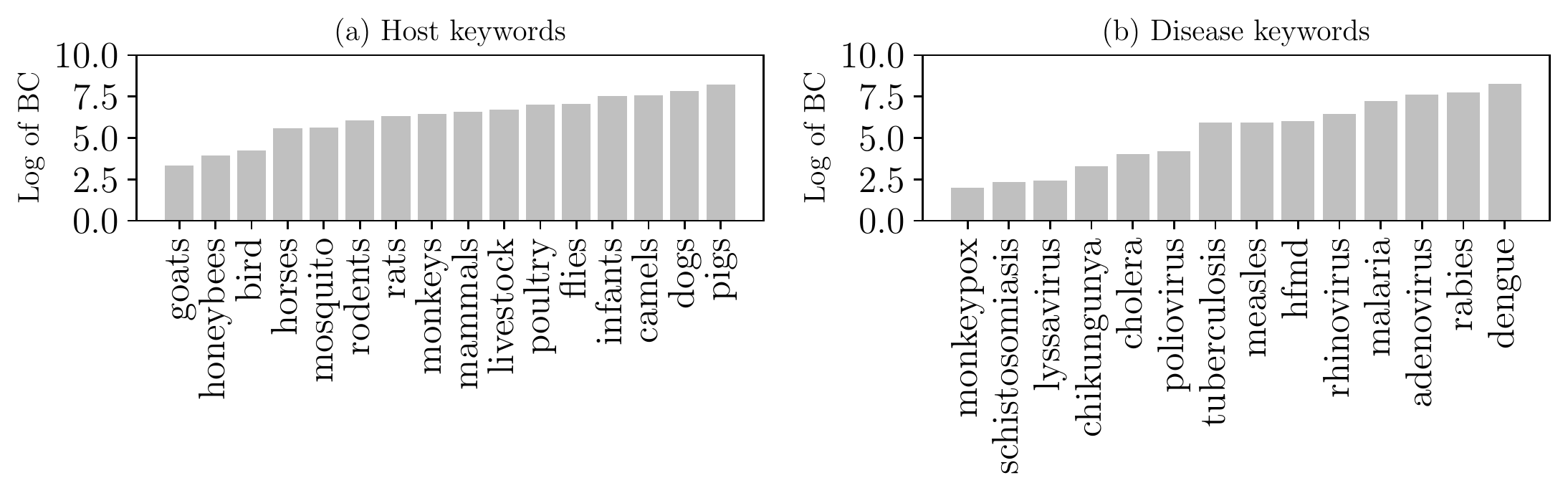}
\caption{Rank ordering of the most important keywords in the \textit{transmission} subgraph that is originally linked to the \textit{coronavirus} node. The subgraph keywords are grouped into (a) host and (b) disease categories. Log of BC is the natural logarithm (log) of betweenness centrality.}
\label{trans01}
 \end{figure}
 
Figure~\ref{trans01} shows the most important keywords in the $transmission$ subgraph based on their BC measurements. Among the pathogen hosts, pigs, dogs, and camels have topped the list. Although bats do not appear in this list of hosts, they do appear later in the gene subgraph. The top diseases in the rank ordering, such as dengue, rabies, malaria, adenovirus, may have some topics common with the study of coronavirus transmission. The rhinovirus related to common cold and lyssavirus often hosted by bats have appeared in the ranking. The adenovirus, ranked in the third position, causes symptoms similar to those of coronavirus infection. 
\begin{table}[t]
\centering
 \caption{Neighboring words in the $drug$ subgraph linked to $coronavirus$.}
\label{tab:drug_keywords}
 \centering
\begin{tabularx}{\textwidth}{l|Xc}
\toprule
Topic &  Connected nodes in the $drug$ subgraph \\
\midrule
 Drug &   'Oseltamivir',    'Amantadine', 'Viagra', 'Tunicamycin', 'Aspirin', 'Paracetamol', ,{\bf'Chloroquine'}, 
 'Meropenem','Troglitazone', 'Bleomycin', 'Dexamethasone',  'Antimoniate', 'Nitazoxanide', 'Niclosamide', 
 'Resveratrol', 'Gemcitabine', 'Rivastigmine', 'Amphotericin', 'Paromomycin', 'Miltefosine', 'Sitamaquine',
  'Berenil', 'Daunorubicin', 'Artemether', 'Buprenorphine', 'Acetaminophen',  'Verdinexor', 'Mizoribine', 
'Amodiaquine',  'Sunitinib', 'Carbapenem', 'Amiodarone', 'Boceprevir', 'Pranobex', 'Dapsone', 'Betalactam', 'Tanespimycin',  'Saracatinib',  'Vildagliptin', 'Ciclosporin', 'Cyclosporin'  \\\midrule
Chemical & 'Glycoprotein', 'Oligonucleotides', 'Glycosylation', 'Cyclophilins', 'Glucocorticoid', 'Acetylcholinesterase', 'Estrogen', 'Cyclooxygenase', 'Glycyrrhizin', 'Hexachlorophene', 'Myricetin', 'Glycosides',  'Hexamethylene', 'Phytomedicines', 'Phytoconstituents', 'Farnesyltransferase', 'Imbricataloic', 'Salicylanilide', 'Quinoxalines', 'Amidohydrolases', 'Enantiomers', 'Ectoenzyme','Isoleucin',  'Hydrochloride',  'Calcium', Glutamate, Phytochemicals', 'Diphosphate', 'Phisohex', 'Monoammonium' \\\midrule
Disease & 'HIV', 'Pneumonia', 'Flavivirus', 'Chikungunya', 'Zikavirus', 'Monkeypox', 'Melanoma','Dengue','Tuberculosis', 'BTCOV',  'Hypertension', 'Phospholipidosis', 'Hematemesis',  'Acinetobacter', 'Baumannii' \\\midrule
Miscellaneous &   'Metabolism', 'Pharmacokinetics', 'Seroprevalence', 'Bactericidal', 'Antifungals',  'Macropinocytosis', 'Toxoplasmosis',   'Syncytia', 'Pyrenees', 'Decoction','Nutraceuticals', 'Chemoprevention', 'Clerodendrum' , 'Molluscicide', 'Myelosuppression', 'Prophylaxis', 'Macrocyclization', \\\bottomrule
 \end{tabularx}
\end{table}

\begin{figure}  [t]
\centering
\includegraphics[scale=0.70]{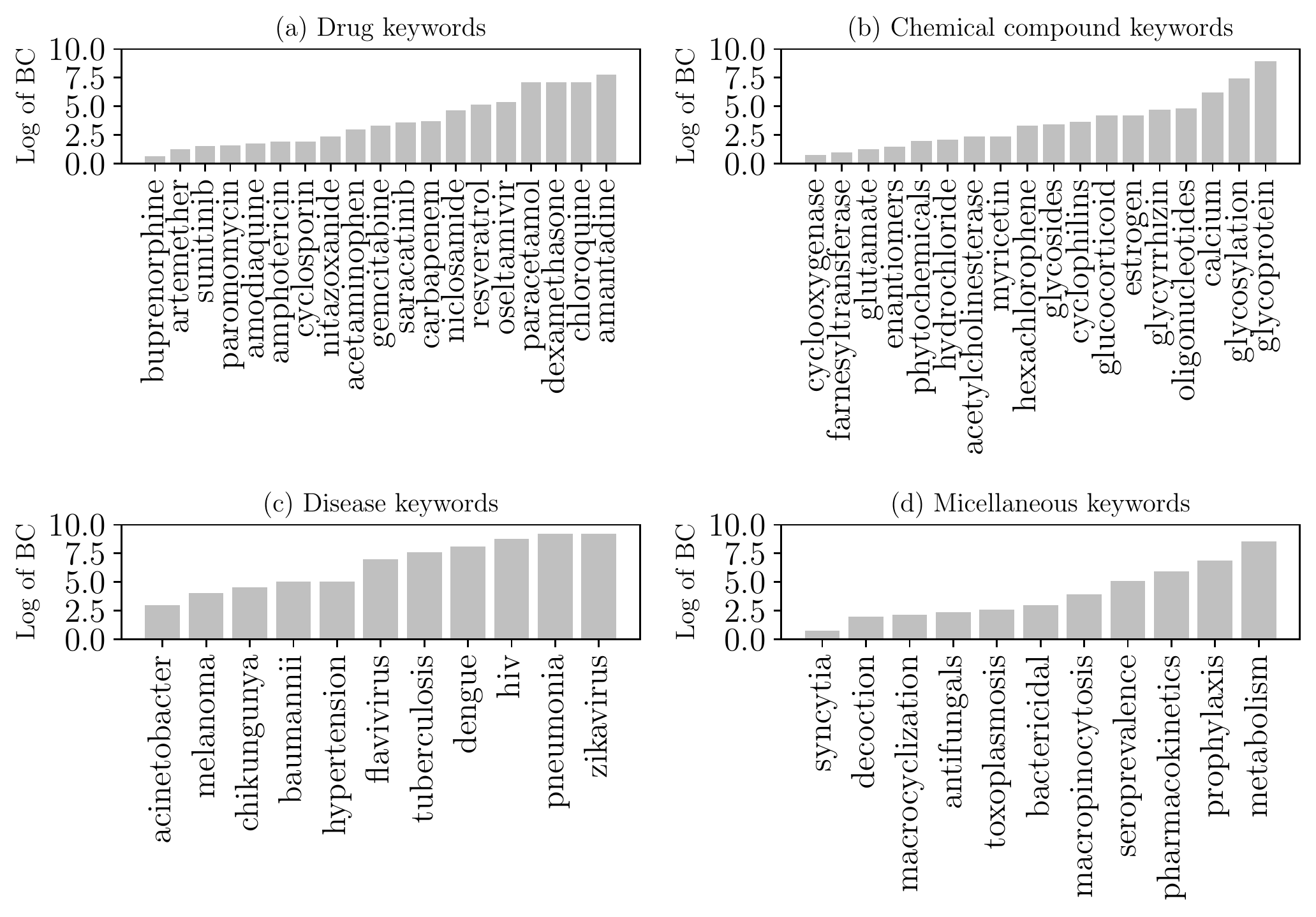}
\caption{Rank ordering of the most influential keywords in the $drugs$ subgraph that is linked to the $coronavirus$ node. The keywords of the subgraph are then grouped into drugs, chemical compounds, diseases, and miscellaneous categories. Log of BC is the natural logarithm (log) of betweenness centrality.}
\label{drugs01}
 \end{figure}

\subsection{Drug Subgraph} 

The study of drugs (pharmacology) plays a big role in the battle against diseases. Table~\ref{tab:drug_keywords} shows the connected words of the $drug$ subgraph that is linked to the $coronavirus$ node. A long list of drugs have appeared in the literature, including Chloroquine. Several antiviral drugs such as 'Verdinexor', 'Oseltamivir', 'Amantadine', 'Pranobex' have appeared in the $drug$ subgraph. Other important drugs that appear in the subgraph are  drugs for cancer treatment (Gemcitabine, Berenil, Daunorubicin, Tanespimycin), anti-fungal drug (Amphotericin), antibiotics (Paromomycin, Carbapenem, Betalactam), immunosuppressive drugs (Mizoribine, Cyclosporin), drugs for malaria (Amodiaquine) and Hepatitis C virus treatment (boceprevir). Many chemical and organic compounds, proteins and enzymes are mentioned in the literature that may have important pharmaceutical properties. Plants (Clerodendrum), plant-based organic compounds (Resveratrol, Glycyrrhizin), and medicine derived from plants (Phytomedicines, Phytochemicals, Phytoconstituents) have appeared in the literature. Bat coronavirus (BTCOV) appears on the list as well. Interestingly, 'Chikungunya', 'Monkeypox', 'Dengue', and 'Tuberculosis' appeared both in the subgraphs of $drug$ and $transmission$. The pathogens of these diseases may have useful links to the study of drug and transmission of coronavirus.

We group the neighboring words within the $drug$ subgraph into four categories: drug, chemical compound, disease, and miscellaneous. Figure~\ref{drugs01} shows the most important keywords under four different categories. Interestingly, antiviral drug $Amantadine$ has the highest rank even before $Chloroquine$. $Amantadine$ is administered to treat symptoms related to flu viruses although it may not be effective for treating illness due to all types of virus strains. $Dexamethasone$ has almost similar BC measurement as $Chloroquine$, which is a drug to aid the immune system in the fight against inflammation. In the disease category of the $drug$ subgraph, zikavirus and pneumonia appear as the top two keywords followed by HIV and dengue. This rank ordering is quite different from that of the $transmission$ subgraph (Figure~\ref{trans01}) where dengue topped the list. There may be common pharmaceutical interests shared by the drugs used for zikavirus and coronavirus treatment. 

In the chemical category, Glycoprotein and a chemical reaction known as Glycosylation have topped the list. Oligonucleotides, single strands of RNA,  have ranked fourth in this category presumably because of their known effect in inhibiting the replication of influenza virus~\citep{Kumar2013}. Interestingly, plant-based compound Glycyrrhizin ranks fifth in the list, which has been used to cure bronchitis, gastritis, and jaundice. There are other proteins and enzymes in the list that will require further discussion by the field experts. The miscellaneous category suggests research studies focusing on 'metabolism', 'prophylaxis' (actions to prevent disease), 'pharmacokinetics' (movement of drugs in the body), and 'seroprevalence' (number of individuals tested positive based on blood serum).

\subsection{Gene Subgraph}
The study of genes is imperative as it distinguishes different virus strains and identifies target proteins for successful discovery of drugs and vaccines. The $gene$ subgraph includes a large number of diseases and pathogens as shown in Table~\ref{tab:gene_keywords}. The list includes Hepatitis viruses (HCV, HBV, HEV), influenza viruses (H1N1, H5N1, H7N9, PR8), other viruses related to Dengue (Denv), Chikungunya (CHIKV), HIV, zikavirus, and malaria. Many of these viruses can share similar genomic structure and replication strategy that can be repurposed in the pathogenic study and drug discovery of novel virus strains such as COV-2. Coronaviruses related to pigs, such as porcine epidemic diarrhea virus (PEDV), porcine hemagglutinating encephalomyelitis virus (PHEV), mouse hepatitis virus (MHV), feline infectious peritonitis (FIP) virus, mice originated human coronavirus (HKU1), alpha- and Betacoronaviruses (mainly infect bats) appear in the subgraph as they belong to the same $coronaviridae$ virus family. In the host list, several species of monkey (Macaque and Cynomolgus) have appeared in addition to pigs and bats. The result is a proof-of-concept that bats are subjects of genetic research related to coronavirus. The discovery of coronavirus strains in hosts like pangolin is a very recent discovery and is still under active research~\citep{Lam2020}. 

\begin{table}[t]
 \caption{Neighboring words in the $gene$ subgraph linked to $coronavirus$.}
\label{tab:gene_keywords}
 \centering
\begin{tabularx}{\textwidth}{l|Xc}
\toprule
Topic &  Connected nodes in the $gene$ subgraph \\
\midrule
 Disease &  'Cancer',  'Zikavirus', 'HCV', 'RSV', 'Pneumonia', 'Influenza','SARS', 'PEDV', 'Dengue', 'HIV', 'IBV', Adenovirus', 'Rabies', 'Hepatitis', 'PRRSV', 'H1N1', 'H5N1', 'CHIKV', 'Rhesus', 'HCOV', 'ARDS', 'Denv', 'Flavivirus', 'HRV', 'Rhinovirus', 'Malaria', 'Rotavirus', 'HMPV', 'Tuberculosis', 'Enterovirus', 'HBV','HEV','COPD', 'Asthma', 'Ebolavirus', 'Herpesvirus'
,'Measles', 'MHV', 'Cytomegalovirus','Sepsis', 'Astrovirus', 'Bronchitis', 'FIP', 'Alphavirus', 'EVB', 'H7N9', 'PVC2', 'CMV', 'Paramyxovirus', 'Betacoronavirus', 'Reovirus', 'VSV', 'Polyomavirus', 'Atherosclerosis','HSV', 'BRSV', 'KSHV', 'CPV', 'Torovirus', 'HKU1', 'Circovirus', 'BRD', 'GBS', 'RRV', 'Cirrhosis', 'Orthoreovirus','Hypoxia', 'Bunyavirus',
'PHEV','Hartmanivirus','PVX', 'Dyskinesia', 'Lassa', 'PR8', 'VWD', 'HPYV6' \\\midrule
Host & 'Macaque', 'Pigs', 'Cats','Bat', 'Mice', 'Swine', 'Poultry', 'Birds', 'Porcine', 'Cattle', 'Camels', 'Cynomolgus', 'Horses', 'Ades' \\\midrule
Biomolecular & 'PCR', 'RNA', 'DNA', 'ACE2', 'NGS', 'Aptamers', 'Metagenomics', 'QPCR', 'SIRNA', 'CD8', 'Ceacam1', 'Haplotypes', 'MIRNA','DPP4', 'KDA', 'STAT3', 'APOD','IFNAR', 'SHRNAS, 'APOE', 'vsiRNAs', 'APOM', 'IFN', 'Glycoprotein', 'Corticosteroids','PRP', 'Macrolide', 'Macrophages', 'Cytokine', 'Interferon', 'Nucleotide', 'Polymerase', 'Demyelination', 'Immunoglobulin', 'Astrocytes', 'Chemokines', 'Myelin','PRF','TH2', 'Kinases', 'Pleiotropic', 'ORF3', 'Phosphoprotein', 'Kegg', 'Caspase', 'Peptidase', 'Eukaryotes', 'Mycoplasma', 'Cryptosporidium', 'Mycobacterium', 'MBL2' \\\bottomrule
 \end{tabularx}
\end{table}

\begin{figure}[t]
\centering
\includegraphics[scale=0.7]{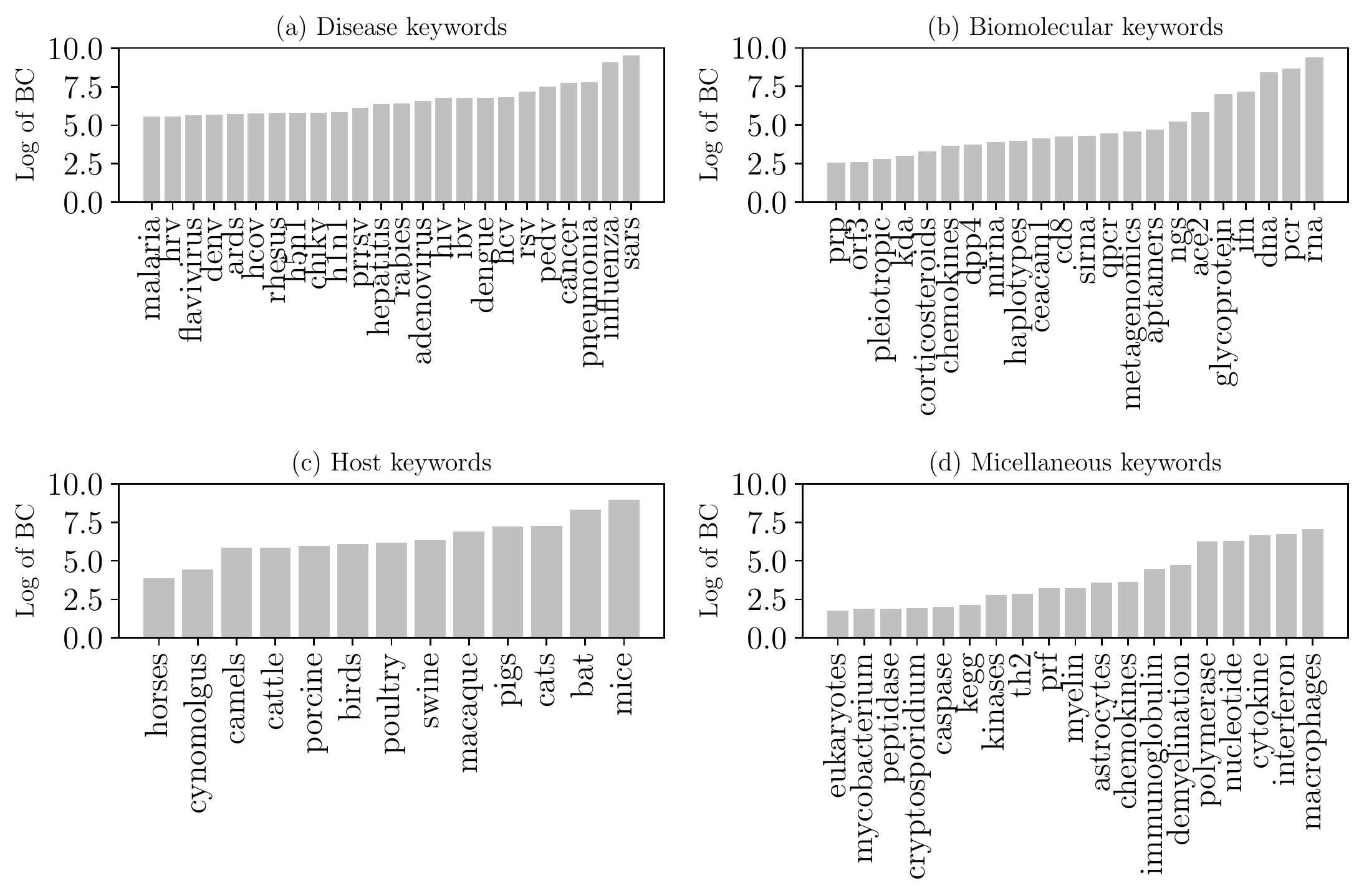}
\caption{Rank ordering of the most influential keywords in the $gene$ subgraph linked to the $coronavirus$ node. The subgraph keywords are grouped into disease, host, biomolecular, and miscellaneous categories. Log of BC is the natural logarithm of betweenness centrality.}
\label{gene01}
 \end{figure}

We further group the neighboring words of the $gene$ subgraph into four categories: host, disease, bio-molecular, and miscellaneous. The biomolecular category includes proteins, cellular mechanisms, drugs, and biochemical compounds. Figure~\ref{gene01} shows the rank ordering of keywords of each category based on BC measurements. Interestingly, cancer appears after SARS, influenza, pneumonia, which suggests a link between the genomic studies of cancer and virus. The porcine epidemic diarrhea virus (PEDV) appears in the fifth position suggesting an important pathogen in the genetic study of flu viruses. In the bimolecular category, polymerase chain reaction (PCR), a mechanism used in molecular biology for testing coronavirus, has ranked second after RNA. The abbreviation of interferon (IFN) appears fourth in this category and second in the miscellaneous keyword category,  which is a set of proteins released by the host in response to a virus attack. In the host category, mice rank the top followed by bat, cat, pig, and a species of monkey (macaque). There are other potentially intriguing keywords in the ranking that will require more space and scope for further discussion.   

\subsection {Summary and Limitations}
We have developed a graph-based model to mine information for COVID-19 research. The major findings are as follows. 1) The proposed graph model is effective in searching informative and connected keywords from a large volume of documents. The BC measurement of individual words can rank their importance that has shown intuitive and informative results. 2) pigs appear in both $transmission$ and $gene$ subgraphs as an important host of pathogens related to the $coronaviridae$ virus family (PEDV and PHEV), 3) in addition to dengue and malaria, zikavirus may have a connection with the development of drugs for flu viruses, 4) our model finds a potential link between genome research of cancer and flu virus, 5) a list of antiviral drugs have been identified where 'Amantadine' ranks above 'Chloroquine' in the treatment of flu virus infection. More importantly, the authors of a recent review on finding therapies for COVID-19~\citep{Li2020} highlight the importance of a) repurposing antiviral agents used in treating HIV, HBV, HCV, and Influenza, b) oligonucleotide and interferon-based therapies, c) glycoprotein as a target for an antiviral agent, d) reference to RNA viruses (Influenza, Ebola, chikungunya, and enterovirus), e) a drug 'Nitazoxanide' to inhibit COV-2 - all of which appear in our subgraph models and rank order. We believe similar other keywords will intrigue the scientists to find repurposing candidates for COVID-19 research.

Despite promising findings, there are several limitations of this study. First, high BC values of top ranked keywords may be partly attributed to their high prevalence in the literature. There is a possibility that keywords in the lower rank order are less prevalent in the literature. However, low-ranked keywords can still be strong candidates for future COVID-19 research. Second, our analysis is limited to article abstracts without incorporating full article texts that can be a richer source of information. Third, we do not use any explicit metric to evaluate our graph model performance and findings since this study is entirely exploratory. Fourth, our graph model includes a broad array of diverse health science literature. A grouping of research articles and subsequent analysis within individual subgroups may yield more focused results. We leave these tasks for our future work.

\section{Conclusions}

This paper demonstrates one of the first studies on mining a large volume of health science literature in search of information for the ongoing battle against COVID-19.  The results presented in tables and figures under different topics effectively summarize the key information contents of the literature with the aid of a graph-based computational model and betweenness centrality based keyword ranking. The proposed model has been able to recommend informative drug types, pathogens, hosts, chemical and biomolecules for COVID-19 research that may not always be evident to the concerned researchers. The findings may be a useful reference guide to provoke new ideas and expedite new directions of research in the fight against the COVID-19 pandemic.

\bibliographystyle{unsrtnat}
\bibliography{covid19_text_final}

\end{document}